# The Impact of Augmented-Reality Head-Mounted Displays on Users' Movement Behavior: An Exploratory Study

Yunlong Wang, Harald Reiterer
HCI Group
University of Konstanz
Konstanz, Germany
{firstname.lastname}@uni-konstanz.de

## ABSTRACT

The augmented-reality head-mounted display (e.g., Microsoft HoloLens) is one of the most innovative technologies in multimedia and human-computer interaction in recent years. Despite the emerging research of its applications on engineering, education, medicines, to name a few, its impact on users' movement behavior is still underexplored. The movement behavior, especially for office workers with sedentary lifestyles, is related to many chronic conditions. Unlike the traditional screens, the augmented-reality head-mounted display (AR-HMD) could enable mobile virtual screens, which might impact on users' movement behavior. In this paper, we present our initial study to explore the impact of AR-HMDs on users' movement behavior. We compared the differences of macro-movements (e.g., sit-stand transitions) and micro-movements (e.g., moving the head) between two experimental modes (i.e., spatial-mapping and tag-along) with a dedicated trivial quiz task using HoloLens. The study reveals interesting findings: strong evidence supports that participants had more head-movements in the tag-along mode where higher simplicity and freedom of moving the virtual screen were given; body position/direction changes show the same effect with moderate evidence, while sit-stand transitions show no difference between the two modes with weak evidence. Our results imply several design considerations and research opportunities for future work on the ergonomics of AR-HMDs in the perspective of health.

## 1 Introduction

Augmented-reality head-mounted displays (AR-HMD, e.g., HoloLens) have been popular in both academia and industry. This multimedia platform enables novel interfaces between users and information, which brings immersive and in-situ user experience. However, the ergonomics, especially the health impacts of using these devices, is still underexplored.

Since the beginning of the PC era, human labor work has been increasingly replaced by machines and computers. Inactive office work becomes many people's lifestyle. However, human's body cannot evolve so fast as the modern industry that many chronic diseases become pervasive. In the short term, static postures that cause much pressure to our spines could contribute to musculoskeletal disorders (e.g., back and neck pain) [14]. In the long term, inactive lifestyles are related to obese [11], type 2 diabetes [27], cardiovascular diseases [37], and even certain cancers (e.g., colon cancer [18]). Recent studies [22] showed that even meeting the well-established physical activity (PA) recommendations (i.e., 150-300 minutes of moderate intensity PA or 75-150 minutes of vigorous intensity PA per week) cannot compensate the bad effect of sedentary behavior (prolonged-sitting) at work.

Therefore, improving individuals' inactive work style is urgent. Both academia and industry have studied many methods - from changing working policies and physical environments [36] to providing dedicated education and reminders [13]. According to Fogg's Behavioral Model (FBM) [10], human behaviors are determined by three factors: ability, motivation, and triggers. Lacking one of these factors will lead to the failure of the target behavior. A recent systematic review [36] illustrated that education plus point-of-prompt PC reminders could be effective in reducing sedentary behavior at work. This mixed intervention strategy enhances two factors in FBM: motivation and triggers. Regarding the third factor, ability, changing workplace settings or adding facilities (e.g., using sit-stand work station or treadmill) could be helpful. One way to increase users' ability is to make the target behavior easier to do, which is the focus of this paper.

Despite the approaches to solving the inactivity problem in office work, we have to ask what causes the inactive lifestyle. The reasons could be multiple: the limited room space, being quiet to avoid interrupting colleagues, focusing on tasks, and so on.

However, we believe the fixed computer screen is a non-negligible factor. To visually obtain the information from the computer, we have to stay close to the screen. Even using a sit-stand work station with multiple screens, users' postures are still restricted by the screen's position and size. Therefore, allowing users to move the screen freely could be a way to encourage their movements during work, which corresponds to the factor of ability in FBM. The AR-HMD (e.g., HoloLens) could enable free movements with the virtual screen in the device.

Therefore, we are eager to explore the potential impact of AR-HMDs on users' movement behavior at office work, which could be applied to improve the screen-based office workers' health. A representative of AR-HMDs is Microsoft HoloLens. It enables the experience of holograms, which is a new visual media with high potentials in human-computer interaction. Four features of HoloLens could meet the requirements of our study design: self-contained (wireless), spatial-awareness, movement-awareness, and augmented reality. In this paper, we will present our exploratory study to answer the following research questions:

1. Will higher simplicity and freedom of moving the virtual screen in HoloLens lead to more movements of the users?
2. What are the main factors affecting users' movements when using the virtual screen in HoloLens?

## 2 Related Work

AR-HMDs combine the advantages of two technologies: augmented reality (AR) and the head-mounted display (HMD). The emerging research of AR-HMDs has enabled applications on education [35], engineering [32], medicine [3], and so on [4,26]. Although increasing work is focused on novel applications of AR-HMDs, the study of their impacts on human health is limited.

Yuan and colleagues [39] systematically reviewed the HMD-caused visual discomfort, indicating that the exposure to HMDs resulted in higher visual discomfort (i.e., simulator sickness and visual strain) compared with exposure to traditional displays such as TV and desktop computer displays. However, this review only covered studies using virtual reality HMDs. There is no systematic review of the visual discomfort impact by AR-HMDs due to the lack of related empirical studies. In a recent paper, Vovk et al. [34] conducted an experiment using HoloLens with 142 subjects in three different industries (i.e., aviation, medical, and space), finding that HoloLens causes only negligible symptoms of simulator sickness across all participants.

Focusing on users' postures, recently, Aromma et al. [1] reported a study evaluating a tablet-based AR system in maintenance work regarding human factors and ergonomics. Their study showed that the users adopted varied kinds of postures, of which some postures may increase the risk of musculoskeletal disorders in the long term. However, this study has two limitations: the selected task only took 20-30 minutes; the maintenance task in the study largely determined the users' selected postures. In other words, the study was too short while the task did not allow much freedom of users' movements and postures. In our study, we address the limitations by designing dedicated virtual-screen based task using HoloLens.

Although the focus of this paper is exploring the potential impact of AR-HMDs on users' movement behavior, the goal behind is to improve users' health in office work. Therefore, we also discuss some related work focused on posture monitoring/guiding and physical activity promotion in office work, which could inspire the ergonomics research of postures and movements using AR-HMDs.

In a recent paper [38], Wu and colleagues proposed ActiveErgo using sensors and an automatically adjustable screen to improve users' sitting postures. The study of ActiveErgo showed that the participants need support to follow the ergonomics guidelines. However, although the right posture is helpful to reduce musculoskeletal pain, prolonged sitting or standing is still detrimental for health. Increasing evidence [2,6] suggests that a dynamic work style – e.g., reducing sitting while increasing sit-stand transitions – is superior to only sitting or standing.

To promote physical activity at work, Probst presented the Active Office [28], where new interaction technologies were designed to enable more diverse movements in office work than the traditional point-and-click interaction. For example, a user controls a PC through the movements of her/his body (e.g., tilting, rotating, bouncing) while sitting on an adjustable office chair with sensors. The proposed technologies in Active Office were focused on the interaction between users and fixed screens, which could still limit users' movements during their office work. However, the idea of integrating more movements to the office work routines also applies in the ergonomics design for AR-HMDs.

## 3 Study Design

### 3.1 Device

Several AR-HMDs are available in the consumer market recently, e.g., Microsoft HoloLens[1], Magic Leap One[2], and HTC Vive Pro[3]. As introduced earlier, HoloLens is a self-contained AR-HMD, which was released by Microsoft in 2016. In our study, we use the HoloLens due to its rich development documents and its large community of AR-HMD research. HoloLens uses the head gaze to control the virtual cursor and the air-tap hand gesture (or a remote clicker) to select the virtual icons. It also has several limitations, e.g., the small field of view and the weight. We considered these limitations when we design the task for our study.

### 3.2 Task

Since we want to explore the effect of the simplicity and freedom of moving the virtual screen in HoloLens on users' movement behavior, we need to design the task that minimizes the impact of other factors. In the review of Yuan and colleagues [39], they discussed the effect of HMDs' optical characteristics (system features), participants' gender (individual characteristics), task duration and content (task characteristics) on users' visual discomfort. We believe these factors will impact on users' movement behavior as well. Therefore, we considered these factors in our study design.

---

[1] https://www.microsoft.com/en-us/hololens
[2] https://www.magicleap.com/magic-leap-one
[3] https://www.vive.com/ca/product/vive-pro/

We used a trivial quiz on a 2D virtual screen as the user task in our study (see Figure 1). The quiz contains 600 hundred questions, which took around one hour to finish. We intently chose interesting questions to decrease the boringness and increase participants' concentration during the study. This task is to simulate a simple office work with moderate cognitive load. The reason for choosing the task based on a 2D virtual screen instead of 3D objects is to avoid the effect of the 3D object on users' movements. To answer one question in the quiz, a user should gaze at the answer button then press the remote clicker. Using the click instead of the air-tap gesture is to avoid the arm fatigue after long-period use.

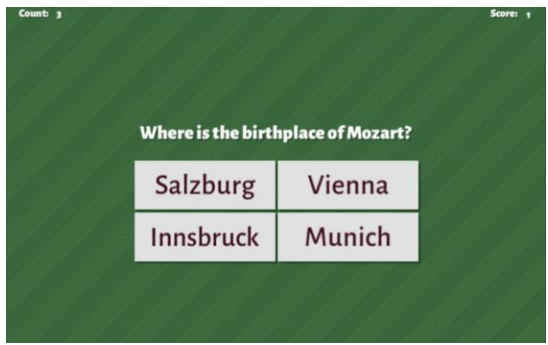

**Figure 1: An question example in the trivial quiz.**

HoloLens has a 30-degree horizontal and 17-degree vertical field of view, which is much smaller than the human eyes' field of view. Therefore, we made the virtual screen a bit smaller than the view area. Thus, it required slight head movement to select the buttons on the screen, which is to minimize the effect of the field of view on users' head movements.

We used two modes – spatial-mapping [24] and tag-along [25] - to provide different levels of simplicity and freedom of moving the virtual screen. In the spatial-mapping mode, users can put the virtual screen to any position on the walls within a given area (around 4 x 4 meters) in a room (see Figure 2). The visual range to the virtual screen is from 0.85 to 3.1 meters away due to the limit of the depth-perception capacity in HoloLens. Differently, in the tag-along mode, the virtual screen automatically follows the user when it is out of the view frustum as the user moves. Otherwise, the virtual screen stays straight to the user's viewpoint at two-meter away. Besides, the spatial-mapping only allows users to attach the virtual screen on the wall in the given room, while the virtual screen has not the limit in the tag-along mode. To conclude, both the modes allow users to control the virtual screen, but the tag-along mode provides more simplicity (automatic following) and freedom (any posture of the virtual screen).

There were a table and an ergonomic chair in the study room. The users could sit, stand, and walk using any postures during the one-hour task. As we aimed to observe users' voluntary behavior, we did not use any indicators or reminders during the study.

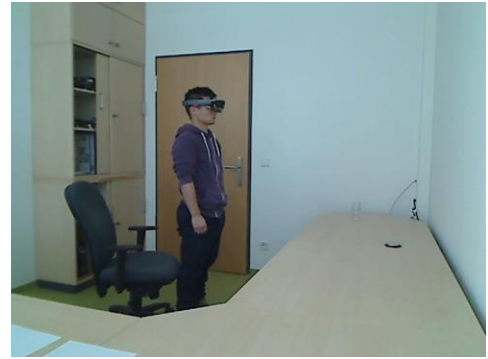

**Figure 2: The physical environment in the study.**

### 3.3 Participants and Procedure

We conducted the study in February 2019. We recruited ten healthy adults (female=5, age=27.2±2.9) from our university using emails and social networks. All the participants had no experience of using HoloLens before the study.

Before the users started the task, we assessed the participants' perceived habit strength using the self-report index of habit strength index [33] and gave them a short introduction of the adverse effect of inactive lifestyle on health. We then told the participants they should focus on the quiz and choose the position and posture as they like during the study. They could see the score on the corner of the virtual screen during the study (See Figure 1). Then the participants signed the informed consent form, which told them the usage of the collected data and that we would give 20 euros to each participant as compensation after the study.

The study consisted of two appointments on two successive days for each participant. We randomly chose five participants to use the spatial-mapping mode at the first appointment and then the tag-along mode at the second appointment. The other five participants used the two modes in the opposite sequence. The questions in the quiz were randomized each time. The cross-over study design was to balance the potential novelty effect of using HoloLens on the two modes. Most of the participants attended the two appointments at the same time on the two successive days, which is to avoid the potential effect of the day time on the participants' behavior. For example, a participant might prefer to sit down in the evening due to tiredness. Two participants were exceptional because of their schedule limitation. Therefore, we counterbalanced their study time for the two modes. After the quiz, we assessed participants' workload using the NASA-TLX questionnaire [17]. The participants were allowed to drop the study or take a break if they feel uncomfortable when wearing the device during the study.

At the end of each appointment, we conducted a semi-structural interview with each participant, which took around 30 minutes. The questions in the interview included:

1. How did you adjust the positions of the virtual screen?
2. How did you choose your position and postures?
3. Which mode do you prefer? And why?
4. Which features would you like to improve or add in the future AR applications on HoloLens?

We used the last two questions only at the end of the second appointment. The interview provided us more insights about the participants' behavior, in addition to objective measurements.

### 3.4 Measurements

We recorded the participants' behavior during the study using two cameras to cover the study room. Besides the videos, we also recorded the postures of HoloLens using the built-in inertial sensors during the study, which corresponded to the participants' head postures.

To quantitatively analyze participants' movement behavior, we borrowed the concepts of macro-movements and micro-movements from Probst's work [8], where she used the macro level and the micro level to categorize users' movements in the office work environment. We developed a scheme (see Table 1) to code the participants' movements from the recorded videos. Sit-stand transitions are the movements a participant sitting in the chair from a standing posture or standing up from a sitting posture. We regarded the posture of leaning to the wall or the table as standing when counting the sit-stand transitions. Body position/direction changes refer to the movements a participant moving her/his body or the chair (while sitting) to another position or direction. Macro-movements require large muscle groups working together. By contrast, micro-movements require less muscle effort, e.g., moving the head, legs, or torso postures. We particularly selected the movement of adjusting HoloLens because it is caused by the discomfort of wearing HoloLens. We have to separate it from other head movements, which includes neck-relaxing movements and intently moving the virtual screen in HoloLens. We did not count the movements of clicking the remote clicker or the air-tap gesture because they were only related to the participants' speed of answering the questions and independent to the mode.

**Table 1: Macro-movements and micro-movements in our coding scheme.**

|  | **Macro-movements** | **Micro-movements** |
|---|---|---|
| **Sub-groups** | Sit-stand transitions; Body position/direction changes. | Adjusting HoloLens; Head movements (excluding adjusting HoloLens); The rest (i.e., arms, legs, and torso movements). |

One author coded the videos when recording the study procedure. After the study, we invited one research assistant (objective coder) to randomly code a subset of the participants' video data [16,23] and then calculated the intra-class correlation (ICC) [31] of the two coders. As we were only interested in the within-subject difference between the two experimental modes, we chose the consistency between the two coders to test the coding reliability. We selected two-hour video data for training the objective coder and five-hour video data for the formal coding. The ICCs was calculated for all the movement categories: sit-stand transitions (ICC = 1.00, p = 0.000), body position/direction changes (ICC = 0.79, p = 0.031), the overall macro-movements (ICC = 0.92, p = 0.005), adjusting HoloLens (ICC = 0.83, p = 0.020), head movements (ICC = 0.81, p = 0.026), micro-movements excluding the head-related ones (ICC = 0.08, p=0.444), and the overall micro-movements (ICC = -0.35, p=0.754). The coding consistency is good (ICC > 0.75) for the most categories. However, the coding for the micro-movements excluding the head-related ones showed poor reliability, which also caused poor reliability of the overall micro-movements. The large variance of the participants' movement habits is the main difficulty of coding the micro-movements, e.g., rotating the chair slightly, shaking legs, swing body slightly. Therefore, we only focus on the measurements with good reliability in data analysis.

## 4　Statistical Analysis

The conventional null-hypothesis significance tests provide little information when the result is not statistically significant – only the alternative hypothesis is tested [7]. Non-significant results might support a null hypothesis over the alternative, or the data are just insensitive. By contrast, Bayes factors [9] compare the extent to what the samples support two hypotheses (e.g., equal or different). Besides, Bayesian methods also allow more principled conclusions from small-n studies of novel techniques in the field of human-computer interaction [20]. Therefore, we use the Bayes factor (**BF**) in addition to the **p-**value [15] and Cohen's **d** [12] to report and interpret the results. We use JASP[4] (Version 0.9.2) for data analysis due to its ability of both the conventional null-hypothesis significance test and the corresponding Bayesian analysis. For readers who are not familiar with Bayes factors, here we provide a short introduction. The definition of the Bayes factor is shown in Formula 1 as below.

$$BF_{01} = \frac{P(H_0 \mid data)}{P(H_1 \mid data)} \text{ or } BF_{10} = \frac{P(H_1 \mid data)}{P(H_0 \mid data)} \quad (1)$$

The Bayes factor is a ratio of the likelihood probabilities. $P(H_0 \mid data)$ is the probability of the null hypothesis given the data, while $P(H_1 \mid data)$ is the probability of the alternative hypothesis given the data. The Bayes factor indicates which hypothesis is more supported by the data. Figure 3 shows the Bayes factor classification and the interpretation we adapted from [8]. The default priors of the alternative hypothesis and the calculation methods for different study design can be found in the work of Rouder and colleagues [29,30].

---

[4] https://jasp-stats.org/

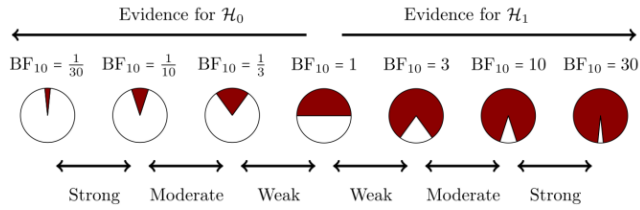

**Figure 3: A graphical representation of a Bayes factor classification and the interpretation, adapted from** [8]**.**

Since we had prior information about the effect of the two modes on users' behavior, we choose the two-sided alternative hypothesis ($H_1$) that the population mean of the difference is not equal to 0. Due to the same reason, we use the default Cauchy distribution ($r = 1/\sqrt{2}$) as the prior when we estimate the effect size. Following the JASP guidelines [8], we also report the median (**M**) and the 95% credible interval (**CI**) of the effect size.

Regarding the result accuracy and reliability, we also checked the normality assumption, the robustness, and the error percentages of calculating the Bayes factors. Only one measurement violates the normality assumption, while all the other results showed low error percentage (<=0.011%) and good robustness. For the one with a deviation from normality, we applied the Wilcoxon signed-rank test and calculated to matched pairs rank biserial correlation (**r**) [21] to show the effect size.

## 5 Results

### 5.1 Macro-Movements

The t-test result shows no significant difference of macro-movement times between the two modes (**p = 0.109**, **Cohen's d = 0.563**). The Bayes factor provides no evidence for both H0 and H1 (**BF$_{10}$ = 0.993, M = 0.455, CI = [-0.137, 1.110]**). However, in the boxplot of the within-subject difference (Figure 4), we see a trend that the macro-movements of the tag-along mode are less than the spatial-mapping mode.

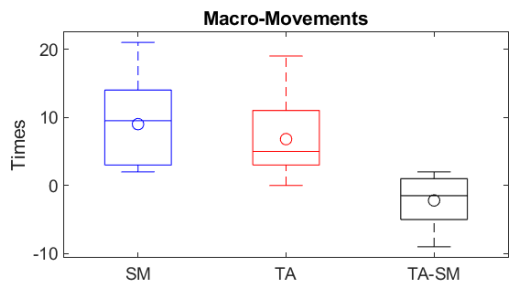

**Figure 4: The boxplot of macro-movements in the spatial-mapping (SM) mode, and the tag-along (TA) mode. The boxplot of within-subject difference (TA-SM) is to show the effect intuitively. The circles in the boxplots refer to the means.**

As macro-movements include sit-stand transitions and the movements of changing body direction/position, we decompose them into two categories for a deeper understanding (see Figure 5). The comparison of sit-stand transitions shows weak evidence for H0 (**BF$_{10}$ = 0.453, M = -0.243, CI = [-0.841, 0.310], p = 0.359, Cohen's d = -0.306**), while the result of body direction/position changes indicates moderate evidence in favor of H1 (**BF$_{10}$ = 3.324, M = 0.704, CI = [0.057, 1.488], p = 0.022, Cohen's d = 0.874**). Therefore, the participants changed their body direction/position when sitting or standing more often in the spatial-mapping mode than the tag-along mode. However, sit-stand transitions tended to be the same. This grouping analysis explains the comparison result of the total macro-movements: the sit-stand transitions data (weakly supporting H0) compensates the moderate evidence for H1 from the body direction/position changes data, resulting in no evidence for both H0 and H1 for the whole macro-movements.

Besides the number of sit-stand transitions, the duration of sitting/standing is also of interest. Because the data of standing duration violate the normality assumption, we use a Wilcoxon signed-rank test to compare the two modes. The result does not suggest a significant effect (**p = 0.183, r = -0.709**). However, from the boxplot (Figure 6) of the within-subject difference of the standing duration, we see most of the difference is positive except one outlier. In other words, all the participants but one stood more in the tag-along mode than the spatial-mapping mode.

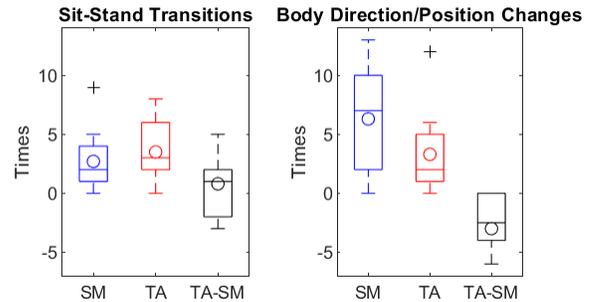

**Figure 5: The boxplot of sit-stand transitions and body direction/position changes in the spatial-mapping (SM) mode, the tag-along (TA) mode, and the within-subject difference (TA-SM).**

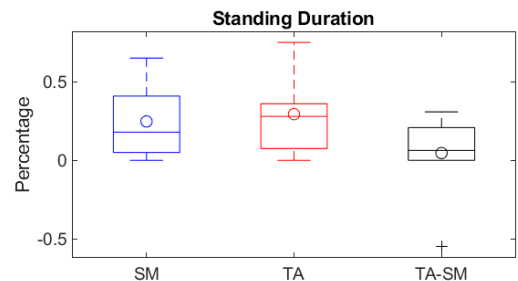

**Figure 6: The boxplot of users' standing duration in the spatial-mapping (SM) mode, the tag-along (TA) mode, and the within-subject difference (TA-SM). The cross in TA-SM refers to the outlier.**

## 5.2 Micro-Movements

Regarding the micro-movements, we investigate two categories: the movements of adjusting HoloLens (AH) and the rest head movements (see Figure 7). The AH involves several movements with hands and the head - caused by the discomfort of wearing HoloLens – showing weak evidence for $H_1$ (**$BF_{10}$ = 1.660, M = 0.560, CI = [-0.044, 1.252], p = 0.054, Cohen's d = 0.700**). The rest head movements indicate strong evidence in favor of $H_1$ (**$BF_{10}$ = 10.900, M = -0.980, CI = [-1.816, -0.154], p = 0.005, Cohen's d = -1.168**) with the other direction. In other words, the participants moved their heads much more frequently in the spatial-mapping mode than the tag-along modes, excluding the movements caused by adjusting HoloLens. Therefore, the comparison of head-related micro-movements shows different trends between the two modes to the overall micro-movements.

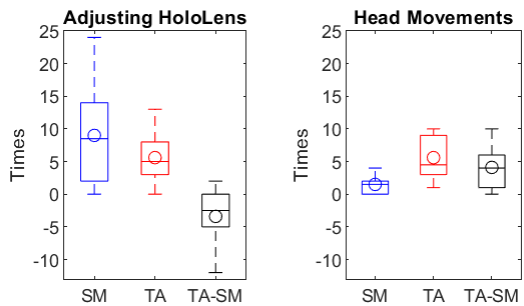

**Figure 7: The boxplot of the numbers of adjusting HoloLens and head movements in the spatial-mapping (SM) mode, the tag-along (TA) mode, and the within-subject difference (TA-SM).**

## 5.3 Head Direction

Besides movements, using right postures when sitting or standing is also critical to prevent musculoskeletal disorders [14], especially the ones caused by new technologies (e.g., the "text neck" [5]). The pitch angle of the head corresponds to the pressure on the cervical vertebrae of the spine. The logged data of the HoloLens postures using the built-in sensors allow us to analyze the pitch angle of the participants quantitatively. We only include seven participants' data because three participants' data of the HoloLens posture were not complete due to technical issues. The Bayes factor shows very weak evidence in favor of $H_1$ (**$BF_{10}$ = 1.092, M = -0.527, CI = [-1.335, 0.163], p = 0.114, Cohen's d = -0.698**). However, the box plot (Figure 8) shows a trend that the pitch angles in the tag-along mode are 6.2 degrees higher than the spatial-mapping mode on average. After checking each participant's data, we found that the difference is mainly caused by the big changes of two participants (#5 and #6 in Figure 9). The medians of the pitch angle are 12.6 and 14.2 degrees in the two modes, which are very close to each other. It should be noticed that the participants changed their head pitch angles several times during the study (see Figure 7). However, the average values indicate that the participants spent more time using the head-up postures.

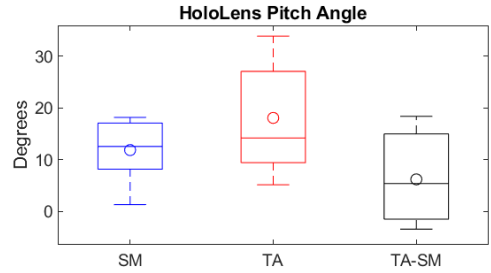

**Figure 8: The boxplot of the numbers of adjusting HoloLens and head movements in the spatial-mapping (SM) mode, the tag-along (TA) mode, and the within-subject difference (TA-SM).**

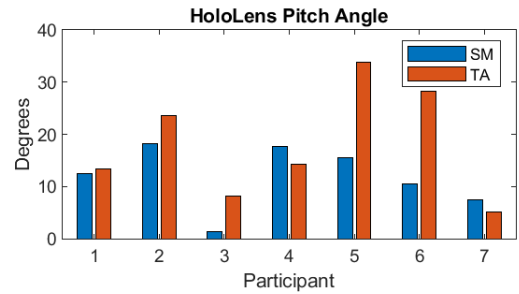

**Figure 9: The bar chart of the pitch angle of HoloLens during the study in the spatial-mapping (SM) mode, and the tag-along (TA) mode. The data from participants #5 and #6 contribute to the most difference on average.**

## 5.4 Workload

The Bayes factors of the workload assessment between the two modes show weak to moderate evidence in favor of $H_0$ (see Figure 10 and Table 2). This result indicates that the participants were under a similar workload during the two modes, as expected. Besides, the scores of the mental demand and the effort suggest that our task caused a moderate workload to the participants.

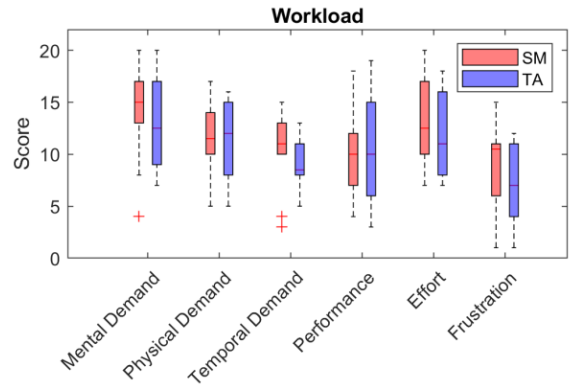

**Figure 10: The boxplots of the participants' scores of the six workload factors in the NASA-TLX questionnaire.**

**Table 2: The Bayes factors and the effect sizes (the medians and the credible intervals) of the workload assessment comparison between the two modes.**

|  | Mental Demand | Physical Demand | Temporal Demand | Performance | Effort | Frustration |
|---|---|---|---|---|---|---|
| **BF$_{10}$** | 0.414 | 0.318 | 0.653 | 0.411 | 0.743 | 0.655 |
| **M** | 0.204 | 0.063 | 0.345 | -0.205 | 0.387 | 0.353 |
| **CI** | [-0.350, 0.786] | [-0.480, 0.617] | [-0.223, 0.973] | [-0.796, 0.340] | [-0.191, 1.010] | [-0.228, 0.969] |

## 6 Discussion

### 6.1 Explanations of Movements

The statistical result of the macro-movements comparison between the two modes moderately supports that the participants performed more movements of changing body position/direction in the spatial-mapping mode. The interview in our study provides us some hints to explain this result: when asked about the factors affecting them choosing postures and adjusting the virtual screen, the common answer of the participants was the physical and visual comfort. Because they could find a comfortable posture to do the task easier in the tag-along mode, they performed fewer movements to adjust their body positions/directions.

Regarding the sit-stand transitions, we see weak evidence in favor of the null hypothesis: no difference between the two modes. Based on the interview, we find that the reasons behind macro-movements are multifaceted. Here we list all the mentioned factors affecting the participants' macro-movements:

1. The musculoskeletal tiredness/discomfort.
   "If getting tired when standing, I will sit down. When feeling uncomfortable using sitting postures, I stand up."
2. The visual tiredness/discomfort.
   "I changed between sitting and standing to find a good posture to optimize visual comfort."
3. The weight of HoloLens.
   "Sometimes I have to sit down and use my hand to hold HoloLens for reducing the weight on my head."
4. Self-reminding of health.
   "I stood up because I thought it's unhealthy to sit for a long time, but not I was uncomfortable when I sat there."
5. Unconsciousness/habits.
   "I think I just sat down after standing for a while because of my habit of sitting."

The head movements excluding the ones of adjusting HoloLens in the tag-along mode were significantly more than the spatial-mapping mode with strong evidence. This result indicates that simplicity and freedom encourage head movements. We also found the reason from the interview: the participants needed to adjust postures to reduce the discomfort of wearing HoloLens and some static postures during the task; the tag-along mode provided an easier way to do so.

Based on the results of macro-movements and micro-movements (head movements), we can answer our research questions:

1. The higher degree of simplicity and freedom of moving the virtual screen leads to more head movements. However, the effect on macro-movements is complex. Weak evidence supports no effect on sit-stand transitions, while moderate evidence shows that the higher degree of simplicity and freedom lead to fewer movements of changing body position/direction.
2. The main factor affecting the participants' movements during our task is the body discomfort including the musculoskeletal and the visual discomfort. The body discomfort might be caused by wearing HoloLens and the participants' static postures. Some participants could remind themselves to stand up after sitting for health, but only occasionally.

Therefore, three factors are related to the participants' movements during the study: the physical and visual discomfort of using HoloLens, the simplicity and freedom of moving the virtual screen, and the participants' ability to remind themselves to move. Comparing to the traditional computer screen, HoloLens provides more movement and posture freedom, but less physical and visual comfort for screen-based tasks.

### 6.2 Reminders

The participants' sit-stand transitions were around three times during the one-hour study session on average in both modes (see Figure 5), while the standing duration accounted for 20-30% of the one-hour session on average (see Figure 6). These numbers seem to indicate a healthy combination of sitting and standing durations. However, some participants still sat for a long time without moving. There were two participants even sitting all the time during the study (see Figure 11).

The score of the self-report habit strength index (5.47±0.65 out of 7) indicates that all the participants had moderate-to-strong sedentary habit strength. Figure 11 shows that the participants with sedentary work styles could still be sedentary even given the freedom of moving the virtual screen. Therefore, reminders for prolonged sitting are also necessary when using AR-HMDs, just as the case when using computer screens.

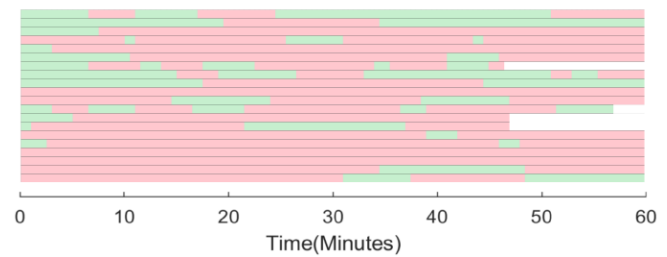

**Figure 11: Each row represents a one-hour study session. The green part refers to standing; the red part refers to sitting; the white part means the participant stopped because of eye fatigue.**

Besides the sitting duration, we also observed several bad postures when the participants wore HoloLens. The postures, which cause an increase of intradiscal pressure, might lead to a higher risk of musculoskeletal disorders over time [14]. Besides some bad postures we can usually see (e.g., slouching in the chair and crossing legs while sitting), we also observed some potentially harmful postures (see Figure 12) that we could hardly use with the traditional screen on the table. Therefore, posture monitoring and reminding are necessary, as we discussed in the section of related work. It could improve the existing AR-HMDs for health promotion by adding these features.

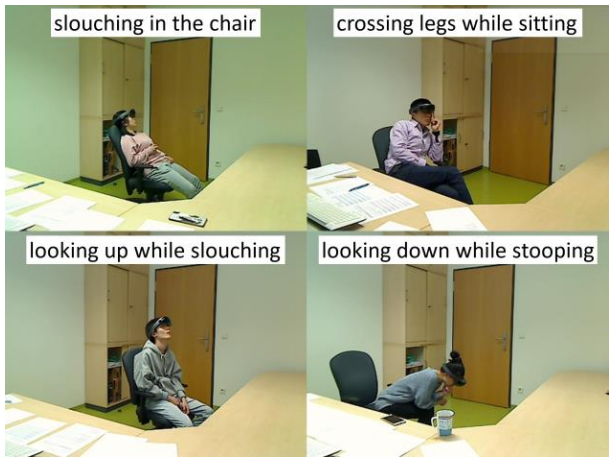

**Figure 12: Selected bad postures of the participants during our study.**

All the participants agreed that a reminder would be helpful because they could easily forget to stand up and use bad postures when focusing on the task. Furthermore, three participants mentioned a prompt reminder might not work for them. They preferred alternative reminders to force them to change postures and move more, e.g., making the virtual screen tilt or blur to indicate bad postures and prolonged sitting.

## 6.3 Limitations and Future Work

This is an initial work to explore the potential impact of AR-HMDs on the movement behavior of office workers. There are several limitations to this work, which could also be research opportunities for future work. First, the user task in this study is simplistic. Real office work might involve more sophisticated user interaction, which should be considered in future work, e.g., browsing webpages or navigating maps.

Second, the wearing comfort of HoloLens might limit the generalization of this work. Due to the disadvantages of weight and the display quality (field of view and resolution) in HoloLens, all participants felt uncomfortable after each one-hour study session. Future work could use other AR-HMDs with a higher degree of wearing comfort. Furthermore, using other AR systems (e.g., RoomAlive [19]) to study users' movement behavior avoiding the effect of discomfort of wearing HMDs also deserves future studies.

Third, we only designed one virtual screen in our study, which did not make full use of the spatial-awareness feature of HoloLens. It would be interesting to study the impact of multi-screen on users' movements.

Lastly, the sample size is relatively small. Despite the moderate to strong evidence in the Bayesian analysis, some evidence is very weak, e.g., the sit-stand transitions and the HoloLens pitch angle. The weak evidence might be improved by using larger sample size.

## 7 Conclusion

Through the exploratory study, we have several interesting findings: (1) moderate evidence supports that body direction/position changes were more in the spatial-mapping mode than the tag-along mode; (2) weak evidence supports that sit-stand transitions and standing durations had no difference between the two modes; (3) the participants adjusted HoloLens more times in the spatial-mapping mode with weak evidence; (4) the participants moved the head more times (excluding the ones related to adjusting HoloLens) in the tag-along mode with strong evidence. The simplicity and freedom of moving the virtual screen encouraged movement behavior, while the discomfort of wearing HoloLens could also make users move to adjust it. All the participants preferred the tag-along mode because they felt it more comfortable, but we also observed bad postures as the side effect of the high simplicity and freedom of moving the virtual screen.

To sum up, this work made the following contributions to the research field of multimedia and human-computer interaction: (1) it is the first work to investigate the potential impact of the augmented-reality head-mounted display (AR-HMD) on users' movement behavior in screen-based office work; (2) inspired by related work, we analyzed the movement behavior through the lenses of macro-movements and micro-movements with categorizing the movements to sub-groups; (3) we used Bayesian method, in addition to the null-hypothesis significant test, to analyze and report the study results; (4) the study results confirmed the effect of the freedom and simplicity (EoFS) of moving the virtual screen in HoloLens on users' movement behavior; (5) besides EoFS, we also found that the discomfort of using HoloLens could partially cause users' movement behavior; (6) based on the findings and the limitations of this work, we provide four research opportunities for future work.

Through this study, we want to show the necessity of studying the ergonomics of AR-HMDs, especially their impact on users' movement behavior. Besides the exciting user experience brought by the new technology, we should also consider the health perspectives when designing applications using AR-HMDs.

## REFERENCES

[1] Susanna Aromaa, Antti Väätänen, Eija Kaasinen, Mikael Uimonen, and Sanni Siltanen. 2018. Human Factors and Ergonomics Evaluation of a Tablet Based Augmented Reality System in Maintenance Work. In *Proceedings of the 22nd International Academic Mindtrek Conference on - Mindtrek '18*, 118–125.https://doi.org/10.1145/3275116.3275125

[2] Jack P. Callaghan, Diana De Carvalho, Kaitlin Gallagher, Thomas Karakolis, and Erika Nelson-Wong. 2015. Is Standing the Solution to Sedentary Office

Work? *Ergonomics in Design* 23, 3: 20–24.https://doi.org/10.1177/1064804615585412
[3] Long Chen, Thomas W Day, Wen Tang, and Nigel W John. 2017. Recent Developments and Future Challenges in Medical Mixed Reality. In *Proceedings of 16th IEEE International Symposium on Mixed and Augmented Reality (ISMAR 2017)*, 123–135.
[4] Pietro Cipresso, Irene Alice Chicchi Giglioli, Mariano Alcañiz Raya, and Giuseppe Riva. 2018. The Past, Present, and Future of Virtual and Augmented Reality Research: A Network and Cluster Analysis of the Literature. *Frontiers in psychology* 9: 2086.https://doi.org/10.3389/fpsyg.2018.02086
[5] Jason M Cuéllar and Todd H Lanman. 2017. "Text neck": an epidemic of the modern era of cell phones? *The spine journal : official journal of the North American Spine Society* 17, 6: 901–902.https://doi.org/10.1016/j.spinee.2017.03.009
[6] Kermit G. Davis and Susan E. Kotowski. 2015. Stand Up and Move; Your Musculoskeletal Health Depends on It. *Ergonomics in Design* 23, 3: 9–13.https://doi.org/10.1177/1064804615588853
[7] Zoltan Dienes. 2014. Using Bayes to get the most out of non-significant results. *Frontiers in psychology* 5: 781.https://doi.org/10.3389/fpsyg.2014.00781
[8] Johnny van Doorn, Don van den Bergh, Udo Bohm, Fabian Dablander, Koen Derks, Tim Draws, Nathan J. Evans, Quentin Frederik Gronau, Max Hinne, Šimon Kucharský, Alexander Ly, Maarten Marsman, Dora Matzke, Akash Raj, Alexandra Sarafoglou, Angelika Stefan, Jan G. Voelkel, and Eric-Jan Wagenmakers. The JASP Guidelines for Conducting and Reporting a Bayesian Analysis. https://doi.org/10.31234/OSF.IO/YQXFR
[9] Bayes Factors, Robert E Kass, and Adrian E Raftery. 1995. Bayes Factors. *Journal of the American Statistical Association* 90, 430: 773–795.
[10] Bj Fogg. 2009. A behavior model for persuasive design. *Proceedings of the 4th International Conference on Persuasive Technology - Persuasive '09*: 1.https://doi.org/10.1145/1541948.1541999
[11] Charles J Fountaine, Meredith Piacentini, and Gary A Liguori. Occupational Sitting and Physical Activity Among University Employees. *International journal of exercise science* 7, 4: 295–301.
[12] Catherine O Fritz, Peter E Morris, Jennifer J Richler, and O Fritz. 2012. Effect Size Estimates: Current Use, Calculations, and Interpretation. *Association* 141, 1: 2–18.https://doi.org/10.1037/a0024338
[13] Benjamin Gardner, Lee Smith, Fabiana Lorencatto, Mark Hamer, and Stuart Jh Biddle. 2015. How to reduce sitting time? A review of behaviour change strategies used in sedentary behaviour reduction interventions among adults. *Health psychology review* 7199, October: 1–24.https://doi.org/10.1080/17437199.2015.1082146
[14] E. Grandjean and W. Hünting. 1977. Ergonomics of posture-Review of various problems of standing and sitting posture. *Applied Ergonomics* 8, 3: 135–140.https://doi.org/10.1016/0003-6870(77)90002-3
[15] Sander Greenland, Stephen J Senn, Kenneth J Rothman, John B Carlin, Charles Poole, Steven N Goodman, and Douglas G Altman. 2016. Statistical tests, P values, confidence intervals, and power: a guide to misinterpretations. *European journal of epidemiology* 31, 4: 337–50.https://doi.org/10.1007/s10654-016-0149-3
[16] Kevin A. Hallgren. 2012. Computing Inter-Rater Reliability for Observational Data: An Overview and Tutorial. *Tutorials in quantitative methods for psychology* 8, 1: 23.
[17] Sandra G. Hart and Lowell E. Staveland. 1988. Development of NASA-TLX (Task Load Index): Results of Empirical and Theoretical Research. *Advances in Psychology* 52: 139–183.https://doi.org/10.1016/S0166-4115(08)62386-9
[18] Regan A. Howard, D. Michal Freedman, Yikyung Park, Albert Hollenbeck, Arthur Schatzkin, and Michael F. Leitzmann. 2008. Physical activity, sedentary behavior, and the risk of colon and rectal cancer in the NIH-AARP Diet and Health Study. *Cancer Causes & Control* 19, 9: 939–953.https://doi.org/10.1007/s10552-008-9159-0
[19] Brett Jones, Lior Shapira, Rajinder Sodhi, Michael Murdock, Ravish Mehra, Hrvoje Benko, Andrew Wilson, Eyal Ofek, Blair MacIntyre, and Nikunj Raghuvanshi. 2014. RoomAlive: Magical Experiences Enabled by Scalable, Adaptive Projector-Camera Units. In *Proceedings of the 27th annual ACM symposium on User interface software and technology - UIST '14*, 637–644.https://doi.org/10.1145/2642918.2647383
[20] Matthew Kay, Gregory L Nelson, and Eric B Hekler. 2016. Researcher-Centered Design of Statistics: Why Bayesian Statistics Better Fit the Culture and Incentives of HCI. In *Proceedings of the 2016 CHI Conference on Human Factors in Computing Systems*, 4521–4532.https://doi.org/10.1145/2858036.2858465
[21] Dave S. Kerby. 2014. The Simple Difference Formula: An Approach to Teaching Nonparametric Correlation. *Comprehensive Psychology* 3: 11.IT.3.1.https://doi.org/10.2466/11.it.3.1
[22] Sara Knaeps, Jan G Bourgois, Ruben Charlier, Evelien Mertens, Johan Lefevre, and Katrien Wijndaele. 2016. Ten-year change in sedentary behaviour, moderate-to-vigorous physical activity, cardiorespiratory fitness and cardiometabolic risk: independent associations and mediation analysis. *British Journal of Sports Medicine*, 1: 1–7.https://doi.org/10.1136/bjsports-2016-096083
[23] Jonathan Lazar, Jinjuan Heidi Feng, and Harry Hochheiser. 2017. *Research methods in human-computer interaction*. Morgan Kaufmann Publishers.
[24] Microsoft. 2018. Spatial mapping design - Mixed Reality | Microsoft Docs. Retrieved April 5, 2019 from https://docs.microsoft.com/en-us/windows/mixed-reality/spatial-mapping-design.
[25] Microsoft. 2018. Billboarding and tag-along - Mixed Reality | Microsoft Docs. Retrieved April 5, 2019 from https://docs.microsoft.com/en-us/windows/mixed-reality/billboarding-and-tag-along.
[26] Robert Patterson, Marc D. Winterbottom, and Byron J. Pierce. 2006. Perceptual Issues in the Use of Head-Mounted Visual Displays. *Human Factors: The Journal of the Human Factors and Ergonomics Society* 48, 3: 555–573.https://doi.org/10.1518/001872006778606877
[27] Hidde P. van der Ploeg, Tien Chey, Rosemary J. Korda, Emily Banks, Adrian Bauman, Cox DR, Pate RR, Brown WJ, Owen N, Proper KI, Tremblay MS, Dunstan DW, Warren TY, Wijndaele K, Stamatakis E, Patel AV, Katzmarzyk PT, van Uffelen JGZ, Grøntved A, Inoue M, Banks E, Craig CL, Brown WJ, Timperio A, Ng S, Ware JE, Lee C, Rockhill B, Chau JY, Hamilton MT, Hamilton MT, and Troiano RP. 2012. Sitting Time and All-Cause Mortality Risk in 222 497 Australian Adults. *Archives of Internal Medicine* 172, 6: 494.https://doi.org/10.1001/archinternmed.2011.2174
[28] Kathrin Probst. 2015. Active Office: Designing for Physical Activity in Digital Workplaces. In *Proceedings of the 14th International Conference on Mobile and Ubiquitous Multimedia (MUM '15)*, 433–438.https://doi.org/10.1145/2836041.2841223
[29] Jeffrey N Rouder, Richard D Morey, Paul L Speckman, and Jordan M Province. 2012. Default Bayes factors for ANOVA designs. *Journal of Mathematical Psychology* 56: 356–374.https://doi.org/10.1016/j.jmp.2012.08.001
[30] Jeffrey Rouder, Paul Speckman, Dongchu Sun, Richard Morey, and Geoffrey Iverson. 2009. Bayesian t tests for accepting and rejecting the null hypothesis. https://doi.org/10.3758/PBR.16.2.225
[31] Patrick E. Shrout and Joseph L. Fleiss. 1979. Intraclass correlations: Uses in assessing rater reliability. *Psychological Bulletin* 86, 2: 420–428.https://doi.org/10.1037/0033-2909.86.2.420
[32] Anna Syberfeldt, Oscar Danielsson, and Patrik Gustavsson. 2017. Augmented Reality Smart Glasses in the Smart Factory: Product Evaluation Guidelines and Review of Available Products. *IEEE Access* 5: 9118–9130.https://doi.org/10.1109/ACCESS.2017.2703952
[33] Bas Verplanken and Sheina Orbell. 2003. Reflections on Past Behavior: A Self-Report Index of Habit Strength. *Journal of Applied Social Psychology* 33, 6: 1313–1330.https://doi.org/10.1111/j.1559-1816.2003.tb01951.x
[34] Alla Vovk, Fridolin Wild, Will Guest, and Timo Kuula. 2018. Simulator Sickness in Augmented Reality Training Using the Microsoft HoloLens. In *Proceedings of the 2018 CHI Conference on Human Factors in Computing Systems*, No. 209.https://doi.org/10.1145/3173574.3173783
[35] Minjuan Wang, Vic Callaghan, Jodi Bernhardt, Kevin White, and Anasol Peña-Rios. 2018. Augmented reality in education and training: pedagogical approaches and illustrative case studies. *Journal of Ambient Intelligence and Humanized Computing* 9, 5: 1391–1402.https://doi.org/10.1007/s12652-017-0547-8
[36] Yunlong Wang, Lingdan Wu, Jan-Philipp Lange, Ahmed Fadhil, and Harald Reiterer. 2018. Persuasive Technology in Reducing Prolonged Sedentary Behavior at Work: A Systematic Review. *Smart Health* 7–8: 19–30.https://doi.org/10.1016/j.smhl.2018.05.002
[37] Darren E R Warburton and Shannon S D Bredin. 2016. Reflections on Physical Activity and Health: What Should We Recommend? *The Canadian journal of cardiology* 32, 4: 495–504.https://doi.org/10.1016/j.cjca.2016.01.024
[38] Yu-Chian Wu, Te-Yen Wu, Paul Taele, Bryan Wang, Jun-You Liu, Ping-Sung Ku, Po-En Lai, and Mike Y Chen. 2018. ActiveErgo: Automatic and Personalized Ergonomics using Self-actuating Furniture. In *The ACM CHI Conference on Human Factors in Computing Systems*, No. 558.https://doi.org/10.1145/3173574.3174132
[39] Yuan J, Mansouri B, Pettey Jh, Ahmed Sf, and Khaderi. 2018. The Visual Effects Associated with Head-Mounted Displays. *Int J Ophthalmol Clin Res* 5, 2: 85.https://doi.org/10.23937/2378-346X/1410085